\documentclass[conference, 10pt]{IEEEtran}
\IEEEoverridecommandlockouts
\usepackage{cite}
\usepackage{amsmath,amssymb,amsfonts}
\usepackage{algorithmic}
\usepackage{graphicx}
\usepackage{textcomp}
\usepackage{xcolor}
\def\BibTeX{{\rm B\kern-.05em{\sc i\kern-.025em b}\kern-.08em
    T\kern-.1667em\lower.7ex\hbox{E}\kern-.125emX}}

\usepackage{amsthm}
\usepackage{algorithm}

\usepackage{multirow}
\usepackage[operators]{cryptocode}

\newcommand{\tecname}{$\mathtt{C^2PI}$}

\usepackage{booktabs}

\begin{document}

\title{$\mathbf{C^2PI}$: An Efficient Crypto-Clear Two-Party Neural Network Private Inference

\thanks{This work was supported in part by the Air Force Research Laboratory (AFRL) and the Defense Advanced Research Projects Agency (DARPA) under agreement number FA8650-18-1-7817.}
}

\author{
    % Authors
    Yuke Zhang\textsuperscript{\rm 1},
    Dake Chen\textsuperscript{\rm 1},
    Souvik Kundu\textsuperscript{\rm 2},
    Haomei Liu\textsuperscript{\rm 1},
    Ruiheng Peng\textsuperscript{\rm 1}, and
    Peter A. Beerel\textsuperscript{\rm 1}
    \\ \small \textsuperscript{\rm 1}Department of Electrical and Computer Engineering, University of Southern California, Los Angeles, CA, USA 
    \\ \small \textsuperscript{\rm 2}Intel Labs, San Diego, USA
    \\ \small \textsuperscript{\rm 1}\{yukezhan, dakechen, haomeili, ruihengp, pabeerel\}@usc.edu, \textsuperscript{\rm 2}souvikk.kundu@intel.com
    \vspace{-3mm}
}
% \affiliations{
%     % Affiliations
%     %\textsuperscript{\rm 1} 
%     Affiliation\\
%     firstAuthor@affiliation1.com, secondAuthor@affilation2.com, thirdAuthor@affiliation1.com
% }

% % % REMOVE THIS: bibentry
% % % This is only needed to show inline citations in the guidelines document. You should not need it and can safely delete it.
% % \usepackage{bibentry}
% % % END REMOVE bibentry

% \begin{document}
\maketitle

\begin{abstract}
Recently, private inference (PI) has addressed the rising concern over data and model privacy in machine learning inference as a service. However, existing PI frameworks suffer from high computational and communication costs due to the expensive multi-party computation (MPC) protocols. Existing literature has developed lighter MPC protocols to yield more efficient PI schemes. We, in contrast, %rethink existing PI frameworks and 
propose to lighten them by introducing an empirically-defined privacy evaluation. To that end, we reformulate the threat model of PI and use inference data privacy attacks (IDPAs) to evaluate data privacy. We then present an enhanced IDPA, named distillation-based inverse-network attack (DINA), for improved privacy evaluation. Finally, we leverage the findings from DINA and propose \tecname, a two-party PI framework presenting an efficient partitioning of the neural network model and requiring only the initial few layers to be performed with MPC protocols. % In particular, we present \tecname, an efficient two-party PI framework where we use DINA to find a boundary layer in a model after which the cryptographic primitives are no longer needed. 
Based on our experimental evaluations, relaxing the formal data privacy guarantees \tecname \ can speed up existing PI frameworks, including Delphi~\cite{delphi2020} and Cheetah~\cite{huang2022cheetah}, up to $2.89\times$ and $3.88\times$ under LAN and WAN settings, respectively, and save up to $2.75\times$ communication costs.

\end{abstract}

% Private inference (PI) has addressed the rising concern over data and model privacy in machine learning inference as a service, however, at the cost of high complexity. In this work, we propose to lighten PI by introducing an empirical-based privacy evaluation that leverages inference data privacy attacks (IDPAs) to evaluate data privacy. We then present an enhanced IDPA, DINA, for improved privacy evaluation. Finally, we leverage the findings from DINA and propose $\mathbf{C^2PI}$, which requires only the initial few layers to be performed with PI. Our experiments show that $\mathbf{C^2PI}$ can speed up existing PI frameworks by more than $3$x.

\section{Introduction}\label{sec:introduction}
With the increasing complexity of deep neural network (DNN) models and their incredible training cost, machine learning inference as a service (MLaaS) has become an inevitable solution saving significant time, cost, and effort in democratizing ML services, even for non-experts \cite{kundu2021analyzing}. 
% However, the increasing concern for privacy brings a challenge that when a client holds private information and a server has a private model, how can inference be performed while protecting each party's privacy? 
However, increasing privacy concerns challenge the inference procedure when a client and a server hold the inference input and network separately and do not want to reveal their private properties to each other. %As shown in Figure~\ref{fig:challenge_mlaas}, 
The client could be a patient with sensitive medical data or a homeowner with private images, and the server could be a hospital system or a commercial company holding proprietary trained DNN models. % IPs.

Private inference (PI) has appeared to address the privacy issue in MLaaS. Existing PI frameworks~\cite{liu2017oblivious, juvekar2018gazelle, delphi2020, gilad2016cryptonets, riazi2019xonn, %demmler2015aby, patra2021aby2, mohassel2018aby3, kumar2020cryptflow, 
rathee2020cryptflow2, tan2021cryptgpu, knott2021crypten, xu2022simc, huang2022cheetah, lehmkuhl2021muse, shen2022abnn2} leverage secret sharing (SS) and multiparty computation (MPC) protocols %, such as \textit{linearly homomorphic encryption (LHE)}~\cite{elgamal1985public, paillier1999public}, and \textit{Garbled Circuits (GC)}~\cite{yao1986generate} 
to enable the participants to jointly perform inference without revealing their input and model parameters to each other. %While most of these works focus on developing secure and efficient protocols, there is no study on which layers in a network should be performed with cryptographic primitives. We try to fill this gap.
Different from prior PI frameworks~\cite{liu2017oblivious, juvekar2018gazelle, delphi2020, gilad2016cryptonets, riazi2019xonn, %demmler2015aby, patra2021aby2, mohassel2018aby3, kumar2020cryptflow, 
rathee2020cryptflow2, tan2021cryptgpu, knott2021crypten, xu2022simc, huang2022cheetah, lehmkuhl2021muse, shen2022abnn2} where data privacy is formally preserved through cryptographic guarantees, we adopt an empirically-defined client data privacy model \cite{li2022ressfl, he2019model, liu2020datamix} to relax PI. In particular, the client's data privacy is defined based on the potential success of inference data privacy attacks (IDPAs)~\cite{he2019model, li2022ressfl}. Specifically, if IDPAs \textbf{cannot} recover the client's input data, the client's data privacy is deemed preserved. IDPAs are typically evaluated through the structural similarity index (SSIM)~\cite{wang2004image} that measures the human perceptual similarity of two images by considering the luminance, contrast, and structure of two images. Users can set an SSIM value (usually $0.3$~\cite{he2019model}) as IDPA's failure threshold, namely, an SSIM below the threshold indicating a failed recovery. The introduction of IDPA-based privacy enables a finer-grained means of quantifying client's data privacy than the Boolean characterization associated with cryptographic protocols.

The inability to recover the client's input with accessible layer outputs implies that the neural network unintentionally preserves the client's input privacy, intuitively due to the irreversibility of the network. 
To investigate the potential for revealing the input from only later layers' activations, we first propose an improved IDPA, i.e., a distillation-based inverse-network attack (DINA) for an improved evaluation as opposed to the baseline IDPAs~\cite{he2019model, li2022ressfl}.

%present a form of inference data privacy attacks (IDPAs)~\cite{he2019model, li2022ressfl}. In particular, to evaluate server's ability to recover client's input, we first propose a distillation-based inverse-network attack (DINA) for an improved evaluation as opposed to the baseline IDPA.

Our attack results indicate that the server indeed often cannot disclose the client's input even if it obtains later layers' outputs. Based on this observation, we propose a novel two-party PI framework, namely, crypto-clear private inference (\tecname), and relax the computational burden of existing PI methods from a new perspective. Specifically, \tecname \  searches for a \textit{boundary layer} in a model, after which the two parties no longer need the cryptographic primitives to preserve the client's SSIM-based data privacy. This allows the server to independently operate on the remaining layers with significantly lower computation and latency. %and share the final results with the client. 
We name the layers before and after the boundary layer as \textit{crypto layers} and \textit{clear layers}, respectively, with the boundary layer as the last crypto layer. Furthermore, we leverage a noise-adding mechanism to further thwart the IDPAs and enhance clients' data privacy. 

The benefit of our \tecname \ is three-folded: \textit{(a) It helps to reduce computational complexity of existing PI schemes. (b) It protects the architecture of the clear layers while existing PI frameworks leak the whole network architecture to the client~\cite{delphi2020,juvekar2018gazelle,huang2022cheetah,rathee2020cryptflow2}.} It is worth mentioning that the carefully designed network architectures are typically considered as intellectual property of network owners~\cite{zhang2021stealing}. \textit{(c) The introduced fine-grained privacy quantification enables users to trade-off PI complexity with the guaranteed level of client's data privacy by tuning the IDPA's failure threshold.} Existing PI frameworks can be considered a special case of \tecname \ where the boundary is at the last layer. 
%where IDPAs are generally %fails even with a success threshold of \textcolor{red}{$0.xx$}.}
%not applicable and the boundary is at the last layer.
%Having later layers not use cryptographic primitive is particularly important, as earlier research showed that these layers attribute to majority of the latency heavy ReLU cost \cite{jha2021deepreduce}. 
We summarize our contributions as follows.
\begin{itemize}
    \item We propose a distillation-based IDPA (DINA), forming an enhanced evaluation of client's data privacy. DINA outperforms existing alternatives~\cite{he2019model, li2022ressfl} by achieving $\mathord{\sim}0.1-0.23$ more structural similarity (SSIM) in image recovery tasks. 

    \item We propose an efficient two-party PI framework, \tecname. To the best of our knowledge, this is the first effort to protect partial neural network architecture, and find the portion of a network that is not necessarily performed with heavy MPC protocols to maintain IDPA-based data privacy. In \tecname, we leverage DINA to find a reliable and conservative boundary between crypto layers and clear layers. Moreover, \tecname \ is orthogonal to recent PI-lightening techniques \cite{ghodsi2021circa,%jha2021deepreduce, 
    cho2022selective, huang2022cheetah, rathee2020cryptflow2, kundu2023making}. Our extensive experimental evaluations show that, \tecname \ can achieve $1.1\times - 1.82\times$ speedup, and save ${\sim}2.5\times$ communication costs for the state-of-the-art two-party PI framework Cheetah~\cite{huang2022cheetah}.
\end{itemize}

\section{Preliminaries}\label{sec:preliminary}
\noindent\textbf{Notations.} Given a pre-trained neural network model $\mathbf{M}$ held by the server %we denote the function of $l$-th layer as $\mathbf{F}_l$. Given 
and an inference input $\mathbf{x}$ held by the client, we denote the output of the first $l$ layers and the inference output as $\mathbf{M}_l(\mathbf{x})$ and $\mathbf{M(x)}$, respectively. As the boundary layer can be after either a linear operation or a ReLU operation, we use decimal $.5$ to denote the ReLU operation. For example, layer $3$ and layer $3.5$ refer to the linear operation and ReLU operation in layer $3$, respectively.

% In this work, one layer contains only linear operations \textbf{or} non-linear operations. Linear operations include convolution, matrix multiplication, and average pooling, whereas non-linear operations include activation functions such as ReLU.

\noindent\textbf{Threat model.} In this work, we follow the semi-honest threat model, where both parties strictly follow the cryptographic protocols, but try to reveal their collaborator's private input by inspecting the information they received. In \tecname, the server is allowed to get the outputs of the boundary and clear layers, from which server will try to recover client's input using IDPAs. 

\noindent\textbf{Inference data privacy attacks (IDPAs).} The inference data privacy was first systematically studied in~\cite{he2019model} for collaborative inference in split learning (SL), where a network $\mathbf{M}$ is split into two parts: $\mathbf{M^1}$ containing the first consecutive layers in $\mathbf{M}$ and $\mathbf{M^2}$ containing the remaining layers. Two participants, edge and cloud, hold $\mathbf{M^1}$ and $\mathbf{M^2}$ respectively. When performing the inference, the edge feeds its input $\mathbf{x}$ into $\mathbf{M^1}$ and sends the result $\mathbf{M^1(x)}$ to the cloud. The cloud then processes $\mathbf{M^2(M^1(x))}$ and shares the inference results with the edge if necessary. In the edge-cloud scenario, the cloud is curious-but-honest trying to recover edge's input $\mathbf{x}$ from $\mathbf{M^1(x)}$ through two kinds of IDPA, i.e., the \textit{maximum likelihood attack (MLA)} and the \textit{inverse-network attack (INA)}~\cite{he2019model}. This threat model is suitable for our client-server scenario in the way that $\mathbf{M^1}$ and $\mathbf{M^2}$ are composed of our crypto layers and clear layers, respectively. Therefore, IDPAs can also be used to evaluate the privacy of our client's input. Assuming that the boundary in $\mathtt{C^2PI}$ is layer $l$, we have $\mathbf{M^1(x)}=\mathbf{M}_l(\mathbf{x})$. Despite the similarities between our client-server scenario and the edge-cloud scenario in %~\cite{he2019model}, 
SL, i.e., (1) we both partition a network into two parts, (2) client and edge need to pass their outputs to server and cloud, respectively, there is a distinction between the two settings, i.e., $\mathbf{M^1}$ is held by the server in the client-server scenario, whereas $\mathbf{M^1}$ is held by the edge in SL. %~\cite{he2019model}. 
Therefore, \tecname \ and SL %~\cite{he2019model} 
address the privacy issue in different situations. In \tecname, the server trains the network and provides service based on its property. In SL, %~\cite{he2019model},
the edge users hold both inference input and a pretrained network but want to move some inference processes to the server because of the limited computation and storage capacities at the edge. 

MLA recovers the input $\mathbf{x}$ by solving an optimization problem $\mathbf{\hat{x}} = \texttt{argmin}_{\mathbf{\hat{x}}}\| \mathbf{M}_l(\mathbf{\hat{x}})-\mathbf{M}_l(\mathbf{x})\|^2_2$ through gradient decent at a target layer $l$. INA constructs an inversion model $\mathbf{M^*}$ with a single architecture, trains the model by taking $\mathbf{M}_l(\mathbf{x'}) (x' \in \text{TraningSet})$ and $\mathbf{x'}$ as the input and output, and recovers the input $\mathbf{x}$ by querying $\mathbf{M^*}$. Conceptually, this new model $\mathbf{M^*}$ approximates the inverse function of first $l$ layers in $\mathbf{M}$. An enhanced INA (EINA) is proposed in~\cite{li2022ressfl} where the inversion model $\mathbf{M^*}$ consists of more powerful residual blocks~\cite{he2016deep}.% for increased strength.

% Evaluation of the attacks: a. peak signal-to-noise ratio (PSNR). PSNR mathematically measures the pixel level recovery quality of the image. b. structural similarity index (SSIM). 

 %Keep in line with~\cite{he2019model}, we assume an SSIM below $0.3$ as a measure of an unsuccessful recovery. In this work, we empirically define the unsuccessful IDPA as a measure of preserved privacy of the inference input.

% We distinguish IDPA from another research hotspot, i.e., training data privacy attack (TDPA)~\cite{zhu2019deep, yin2021see}. IDPA leverages the intermediate layer outputs in collaborative inference to recover the inference inputs. In contrast, most TDPAs leverage gradient in collaborative training to recover the training data. Moreover, the gradient information is not accessible in collaborative inference. For further details on privacy attacks readers may refer to ~\cite{he2019model}.

% \begin{Definition}
% The privacy of the inference input can be quantified by the results of inference data privacy attacks. Specifically, if the SSIM between the recovered image and the original image is less than 0.3, the privacy of the inference input is preserved.
% \end{Definition}

\noindent \textbf{Private Inference schemes.} 
A series of works have been proposed for two-party PI %over the semi-honest threat model 
in the past few years. CryptoNets~\cite{gilad2016cryptonets} proposes a Homomorphic encryption (HE)-only approach, which requires changing the network structure and retraining the network with HE-friendly activation functions, such as the square function. MiniONN~\cite{liu2017oblivious} first combines SS, Garbled Circuits (GCs) and linearly homomorphic encryption (LHE), to perform activation functions like ReLU without changing the network structure. Gazelle~\cite{juvekar2018gazelle} optimizes the LHE techniques to speed up the inference runtime. XONN~\cite{riazi2019xonn} leverages only GCs for binarized neural networks. Delphi~\cite{delphi2020} splits PI into offline (preprocessing) and online phases and moves most heavy cryptographic computations offline. %Another line of PI leverages ABY share-conversion techniques to convert among arithmetic shares, Boolean shares, and GC shares~\cite{demmler2015aby, patra2021aby2, mohassel2018aby3}. CrypTFlow~\cite{kumar2020cryptflow} and 
CrypTFlow2~\cite{rathee2020cryptflow2} proposes more efficient protocols for non-linear layers and division, yielding more than $20\times$ faster PI than Delphi. Cheetah~\cite{huang2022cheetah}, the state-of-the-art two-party PI strategy, presents an oblivious transfer (OT)-based protocol for non-linear operations and achieves $2\times-5\times$ speedup than CrypTFlow2.
%CryptGPU~\cite{tan2021cryptgpu} and CrypTen~\cite{knott2021crypten} develop open libraries for building commonly-used cryptographic primitives, such as SS and LHE, with PyTorch on GPU. 
Recently, PI protocols have also been proposed over the malicious client~\cite{lehmkuhl2021muse, chandran2021simc, xu2022simc}.

\noindent\textbf{Privacy goals.} In \tecname, the client can only learn server's network architecture of the crypto layers, and the output of the inference. All parameters of server's network should be hidden. On the other hand, the server can learn only the outputs of the boundary and clear layers, as well as the inference output. %all other information about the 
Client's private input should not be revealed during operations of MPC protocols nor recovered by IDPAs. While \tecname \ achieves cryptographically formal guarantee on server's model parameter privacy, it targets the empirical SSIM-based protection on client's input data.
Physical attacks such as side-channel attacks (SCAs)~\cite{maji2021leaky, zhang2021stealing} are out of the scope. However, the defenses against SCAs are possibly to be layered on top of our technique. 
\vspace{-1mm}
\begin{figure}[t]
    \centering
    \includegraphics[width=0.8\columnwidth]{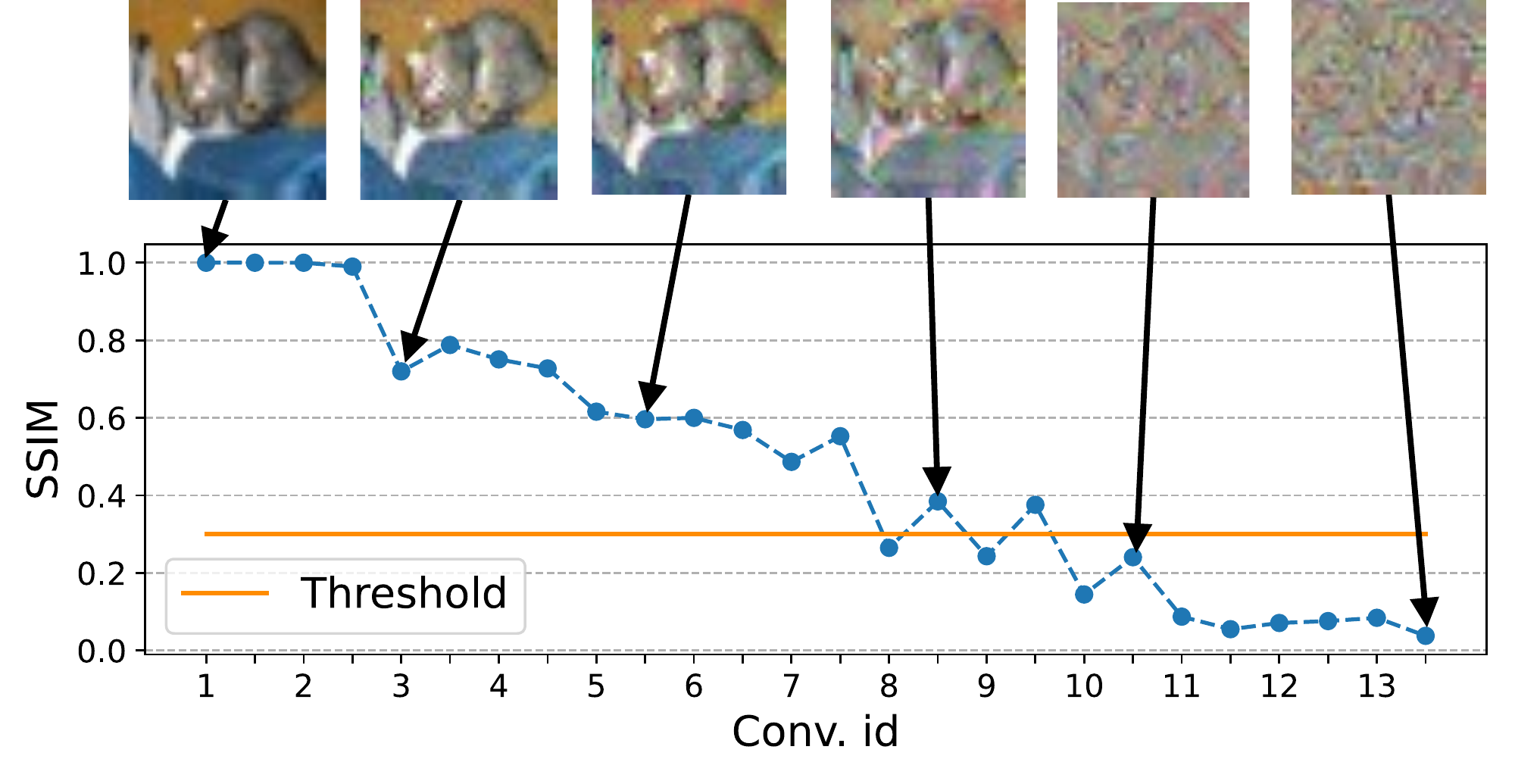}
    \vspace{-4mm}
    \caption{MLA result on an image from CIFAR-10. When SSIM is below the threshold (0.3), the recovered image is difficult to identify.}
    \label{fig:case_study}
\end{figure}

\begin{figure}[t]
    \centering
    \includegraphics[width=0.95\columnwidth]{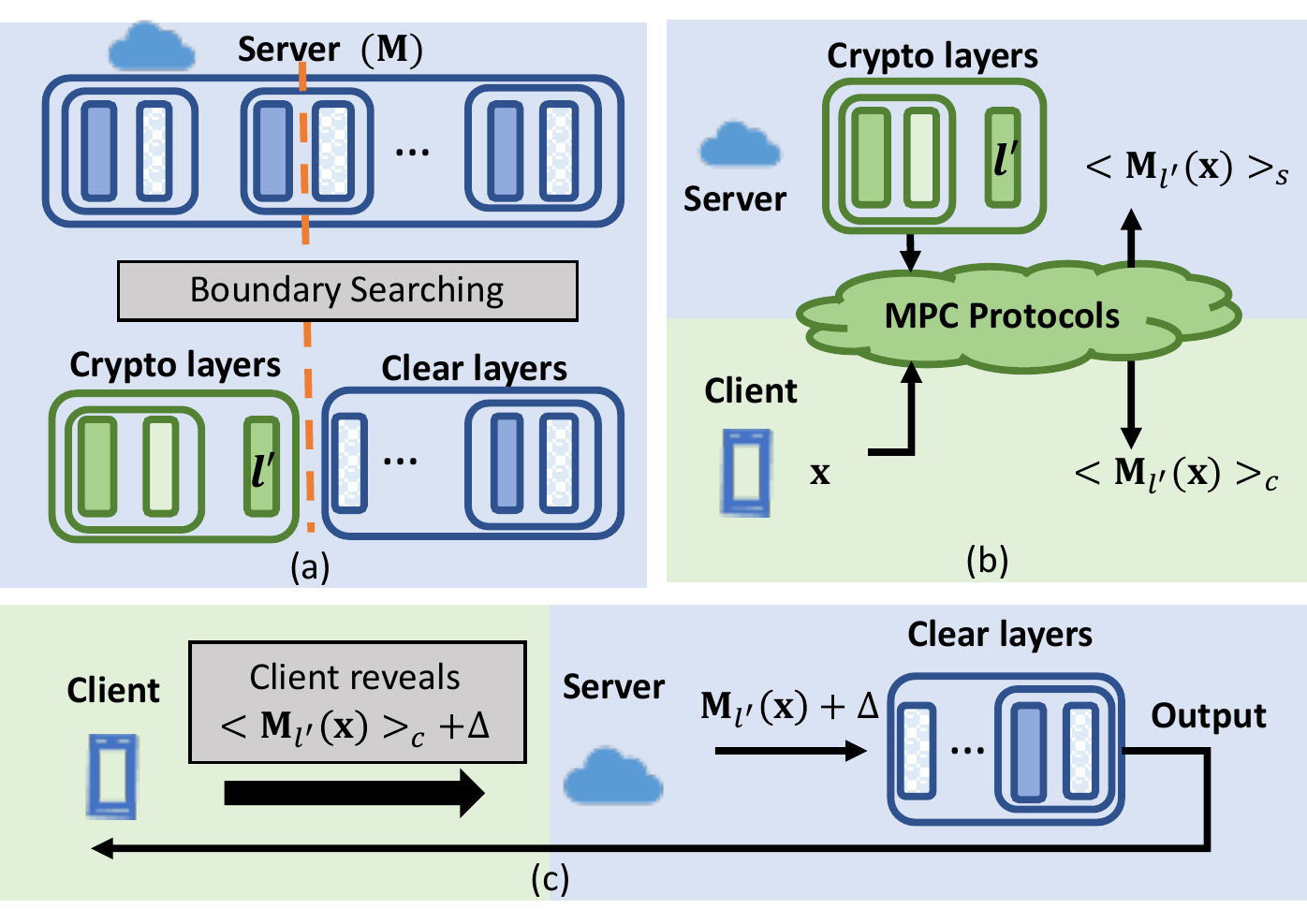}
        \vspace{-4mm}
    \caption{Framework of \tecname: (a) Server identifies the boundary layer $l'$. (b) Operations in crypto layers are performed with MPC protocols (c) Client reveals its share to server. Then, server independently performs the operations in clear layers and shares the final result with the client.}
    \label{fig:c2pi_framework}
    \vspace{-4mm}
\end{figure}

\section{\tecname: Crypto-Clear Private Inference}
We now show that the network naturally helps to hide the original input when the inference procedure enters into deeper layers. 
% This observation implies an opportunity to reduce the cryptographic primitives in the deeper layers.

\noindent\textbf{Case study.} We pretend to be the curious server and use MLA at each layer of a VGG16~\cite{simonyan2014very} model, denoted as $\mathbf{M}$, to reconstruct the client's input $\mathbf{x}$, which is one image from CIFAR-10~\cite{krizhevsky2012imagenet}. When we target layer $l^*$, we assume that we have the intermediate outputs at layers $l\geq l^*$, i.e., $\mathbf{M}_{l}(\mathbf{x}) (l\geq l^*)$, and we do not have the outputs at layers $l<l^*$, i.e., $\mathbf{M}_{l}(\mathbf{x}) (l<l^*)$. The attack results are presented in Figure~\ref{fig:case_study} where SSIM becomes less than $0.3$ after layer $10$. This experiment indicates that server cannot recover client's input from $\mathbf{M}_l(\mathbf{x}) (l\geq10)$ through MLA. Therefore, server could potentially be allowed to have layer outputs after layer $10$ for a lighter PI procedure without violating the client's data privacy.

% \revised{
% \Proposition Explain why MLA fails }

\subsection{\tecname \ Framework}
Figure~\ref{fig:c2pi_framework} shows the \tecname \  framework. Firstly, the server searches for the boundary between the crypto and the clear layers following Algorithm \ref{alg:search_boundary}. In the semi-honest threat model, a curious server will not deviate from this algorithm, %cheat
%deceive in this step, 
and a third-party notary organization can be involved to ensure honesty. Then, server and client jointly perform the operations in crypto layers with a chosen two-party PI method, e.g., Delphi~\cite{delphi2020} or Cheetah~\cite{huang2022cheetah}. At the end of PI, client and server each hold an additive share of the boundary layer's output. Client then adds uniform-distributed noise to its share and reveals the noised share to the server~\cite{pham2022binarizing, titcombe2021practical}. Server searches the maximum noise magnitude in Algorithm~\ref{alg:search_boundary} that yields an acceptable inference accuracy and deliver this to client before performing PI. After summing up the two shares, server performs the operations in clear layers on its own and reveals the inference output to the client at the end of \tecname.

% \noindent\textbf{Boundary searching.} 
The boundary searching algorithm (Algorithm~\ref{alg:search_boundary}) contains two phases. During the first phase (line 1 to line 6), the server sweeps the layers from tail to the head of the model and applies IDPA to recover the inference input. Phase 1 terminates at layer $l'$ at which IDPA begins to succeed in recovery. The potential boundary layer, i.e., the last crypto layer, is layer $l'+1$. Then, in the second phase (line 7 - line 11), server checks the accuracy by assuming the input of layer $l'+2$ becomes the noised input. If the accuracy is above an agreed threshold $\delta$, $l'+1$ is returned as the boundary layer. Otherwise, the server checks the layers after $l'+1$ in sequence until obtaining a satisfying accuracy.

As the operations in crypto layers are performed with existing PI schemes, the data privacy of both parties at these layers is formally proved. Recalling that client's data privacy at clear layers depends on the quality of the IDPA, %Specifically, if the server cannot recover the client's input through a powerful IDPA, the data privacy is preserved in the clear layers. 
we propose a distillation-based inverse-network attack (DINA) to evaluate client's data privacy at clear layers and find the boundary layer. 

\begin{algorithm}[tb]
\caption{Crypto-Clear Boundary Searching}
\label{alg:search_boundary}
\textbf{Input}: A network model $\mathbf{M}$, the number of layers $n$, accuracy threshold $\delta$, noise magnitude $\lambda$, ssim-threshold $\sigma$\\
% \textbf{Parameter}: Optional list of parameters\\
\textbf{Output}: Boundary layer id $l'$
\begin{algorithmic}[1] %[1] enables line numbers

\STATE $l'=n-1$
\STATE $avg\_ssim = {\rm IDPA}(l')$
\WHILE{$avg\_ssim < \sigma$}
\STATE $l'=l'-1$
\STATE $avg\_ssim = {\rm IDPA}(l')$
\ENDWHILE

\STATE $l'=l'+1$

\STATE $n\_acc = accuracy(l', \lambda)$
\WHILE{$n\_acc < \delta$}
\STATE $l'=l'+1$
\STATE $n\_acc = accuracy(l', \lambda)$
\ENDWHILE

\STATE \textbf{return} $l'$
\end{algorithmic}
\end{algorithm}

% As pointed out in~\cite{pham2022binarizing, titcombe2021practical}, adding noise is a common defense against IDPAs.

% \textcolor{gray}{
% Assuming DINA is performed at layer $l'$, the server first constructs a network $\tilde{M_{l'}}$ to approximate the inverse function of the first $l'$ layers in the original model. In distillation-based model inversion attack, $\tilde{M_{l'}}$ is composed by the basicblocks in ResNet~\cite{}. (why basicblocks.) Each basicblock contains residue functions and shortcut functions. The construction topology is that one basicblock corresponds to a block, including a convolutional layer, a batch normalization layer (if applicable) and a ReLU layer, in $M_{l'}$. (what to do with an incomplete block.) Then, the server trains this model $\tilde{M_{l'}}$ with $M_{l'}(x)$ as the input and $x$ as the output. To further enhance the performance of the attack, we introduce distillation points to guide the training. As shown in Figure~\ref{}, the output of each block ...
% The loss function of our distillation-based attack is}
% \begin{equation}
%     \mathcal{L}= \lambda_0()+\lambda_1()+...+\lambda_n()
%     \label{eq:loss_function}
% \end{equation}

\subsection{Distillation-Based Inverse Network Attack (DINA)} 
% \noindent\textbf{Distillation-based model inversion attack.} 

% In the first phase of boundary searching algorithm (Algorithm~\ref{alg:search_boundary}), the proposed DINA aims at developing a model $\tilde{M}$ that recovers the inference input $x$, the input to the model is the output of the layer $l'$: $M_{l'}(x)$. Conceptually, the model approximates the inverse function of layers before $l'$. Keeping in line with~\cite{he2019model}, an SSIM below $0.3$ indicates an unsuccessful recovery.

Despite EINA~\cite{li2022ressfl} increasing the complexity of the inversion model to enhance its inverse ability, it does not take full advantage that server has access to the intermediate layer outputs of its own model, which can be used to guide the training of the inversion model. Therefore, we introduce distillation points in DINA to help the inversion model better approximate the target inverse function.

\begin{figure}[t]
    \centering
    \includegraphics[width=0.82\columnwidth]{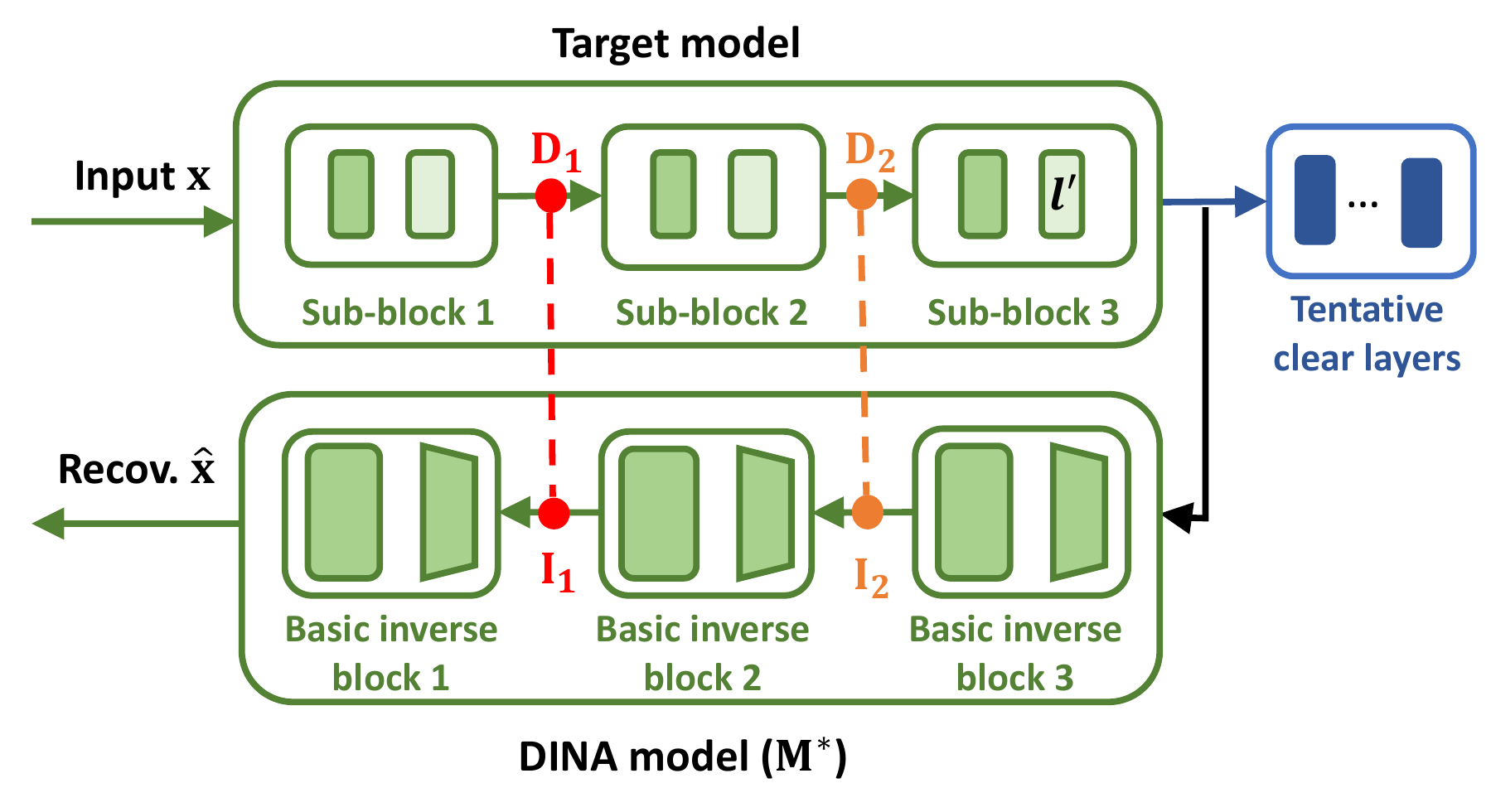}
        \vspace{-5mm}
    \caption{Model architecture of DINA}
    \label{fig:DINA_model}
    \vspace{-5mm}
\end{figure}

Figure~\ref{fig:DINA_model} presents the model architecture in DINA, which is composed of a sequence of \textit{basic inverse blocks}. Each basic inverse block consists of a ResNet basic block~\cite{he2016deep} and a dilated convolution layer. Since the ReLU layer significantly affects the attack, we partition the tentative crypto layers before $l'$ into sub-blocks that end with a ReLU layer, namely, each sub-block only contains one ReLU layer. The proposed attack then uses a basic inverse block to recover the input of one sub-block, as shown in Figure~\ref{fig:DINA_model}, each basic inverse block approximates the inverse function of the sub-block above it.

To better train each basic inverse block, DINA selects middle points between sub-blocks as distillation points \cite{kundu2021attentionlite} and 
applies a fine-grained distillation approach that
optimizes the distance between the output of each basic inverse block and the feature map on the corresponding distillation point. The distances are incorporated into a new loss function:
%and the new loss function is shown below:
   \vspace{-1mm}
 \begin{equation}
    % \mathcal{L}_{DINA}=\sum_{j=1}^{N} \alpha_j \|\mathbf{D}_j-\mathbf{IO}_j\|_2^2+\alpha_0\|\mathbf{x}- \mathbf{M}^*(\mathbf{M}_{l'}(\mathbf{x}))\|_2^2
    \mathcal{L}_{DINA}=\sum_{j=1}^{N} \alpha_j \|\mathbf{D}_j-\mathbf{I}_j\|_2^2+\alpha_0\|\mathbf{x}- \mathbf{\hat{x}}\|_2^2
\label{eq:dina_loss}    
    \vspace{-1mm}
\end{equation}

where the first term is the weighted sum of distance terms at distillation points, $\alpha_j$ is the coefficient that controls the weight of the distance at distillation point $j$, $\mathbf{D}_j$ denotes the feature map at distillation point $j$ in the target model, $\mathbf{I}_j$ is the %intermediate output from 
input of basic inverse block $j$ in DINA model, and $N$ represents the total number of selected distillation points. The second term is the distance between the inference input $\mathbf{x}$ and the output $\mathbf{\hat{x}}$ from DINA model. 

To assist a distillation point in providing effective guidance on its nearest basic inverse block, the attack applies monotonously increasing coefficients $\alpha_j$ from the output to input of DINA model: $\alpha_0<\alpha_1<\alpha_2 ...<\alpha_N$, this ensures that each basic inverse block obtains the most guidance from its nearest distillation point. In the example shown in Figure~\ref{fig:DINA_model}, there are two distillation points, colored in red and orange, respectively. Although the losses at the output of the DINA model and both distillation points contribute to optimizing parameters in the basic inverse block 3, the loss at the orange distillation point has the largest impact due to the monotonously increasing coefficients.

\section{Evaluations}
Our experiments are conducted on AlexNet~\cite{krizhevsky2012imagenet} and variants of VGG16/19~\cite{simonyan2014very}. %and ResNet34~\cite{he2016deep} models
% \footnote{We use variant models with reduced channel size. The models are available at https://anonymous.4open.science/r/c2pi-256C}. 
We train these models on CIFAR-10 and CIFAR-100~\cite{krizhevsky2009learning} with an Nvidia A100 GPU.

\begin{figure}[t]
    \centering
    \includegraphics[width=\columnwidth]{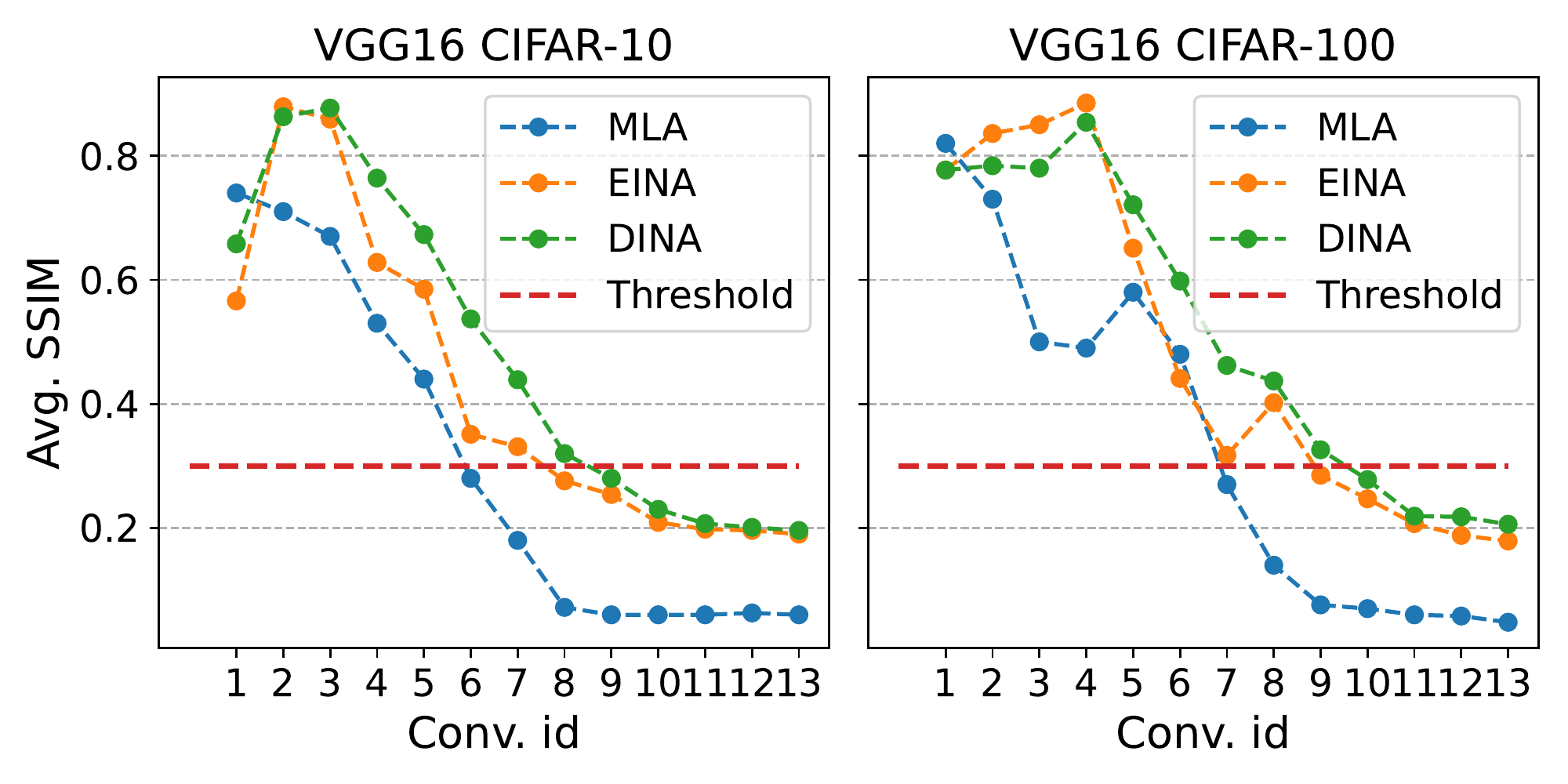}
        \vspace{-8mm}
    \caption{Comparison of IDPAs including MLA, EINA, and DINA.}
    \label{fig:exp_results_attack_comparison}
    \vspace{-4mm}
\end{figure}

% \noindent\textbf{Experiment settings.} 

\subsection{Comparison of IDPAs} 
We apply MLA~\cite{he2019model}, EINA~\cite{li2022ressfl}, and DINA on each layer of VGG16 to recover images in CIFAR-10 and CIFAR-100 datasets. When targeting layer $l$, MLA solves $\mathbf{\hat{x}} = argmin_\mathbf{\hat{x}}\| \mathbf{M}_l(\mathbf{\hat{x}})-\mathbf{M}_l(\mathbf{x}) \|^2_2$ through gradient descent with $10000$ iterations and randomly initialized $\mathbf{\hat{x}}$.
% The denoising process in MLA-De is implemented with the \textit{fastNlMeansDenoisingColored} function in OpenCV library~\cite{}. 
 In EINA, we construct an inversion network $\mathbf{M^*}$ with residual basic blocks~\cite{he2016deep} and train it using the loss function of $\mathcal{L}_{EINA}=\|\mathbf{x}-\mathbf{M^*}(\mathbf{M}_l(\mathbf{x}))\|_2^2$ and stochastic gradient descent optimizer. In DINA, we introduce distillation points and train $\mathbf{M^*}$ with the loss function in~\eqref{eq:dina_loss}. The coefficients in our training are monotonously increasing as $\alpha_0=1, \alpha_1=3, \alpha_{j}=2*\alpha_{j-1} (j\geq 2)$. Both training processes %experience 3 epochs 
 are with a 0.001 learning rate. After training the model $\mathbf{M^*}$, we run the inference over 1000 images from each dataset and evaluate the recovery ability. The noise magnitude and the IDPA failure threshold for this experiment are $0.1$ and $0.3$, respectively.
 
Attack results are presented in Figure~\ref{fig:exp_results_attack_comparison}, where DINA achieves $0.229$ and $0.205$ more average SSIM than MLA at layer $7$ on CIFAR-10 and CIFAR-100, respectively. DINA also presents $0.108$ and $0.145$ more SSIM than EINA at layer $7$ on CIFAR-10 and CIFAR-100.
%%generally achieves a higher average SSIM than MLA and EINA when going into deeper layers. 

Recalling that Algorithm~\ref{alg:search_boundary} first searches for a potential boundary layer after which IDPA begins to fail. MLA, EINA, and DINA return layers $7.5$, $8.5$, and $9$ as the potential boundary layer for CIFAR-10, respectively, and layers $7.5$, $9.5$, and $10$ for CIFAR-100, respectively. Therefore, DINA finds a more conservative boundary than MLA and EINA.

% Attack results are presented in Figure~\ref{fig:exp_results_attack_comparison}, where our DINA performs better than the other three attacks by achieving 0.259 and 0.192 more average SSIM than MLA, 0.121 and 0.032 more than MLA-De, 0.108 and 0.145 more than EINA at layer 22 on CIFAR-10 and CIFAR-100, respectively. As the boundary searching algorithm starts from the last layer, the results at later layers are more important for \tecname. Results in Figure~\ref{fig:exp_results_attack_comparison} indicate that our DINA can recover higher-quality images than other candidates at middle and later layers. Therefore, the boundary searching algorithm will return a more reliable and conservative boundary with our DINA. 

\begin{figure}[t]
    \centering
    \includegraphics[width=0.85\columnwidth]{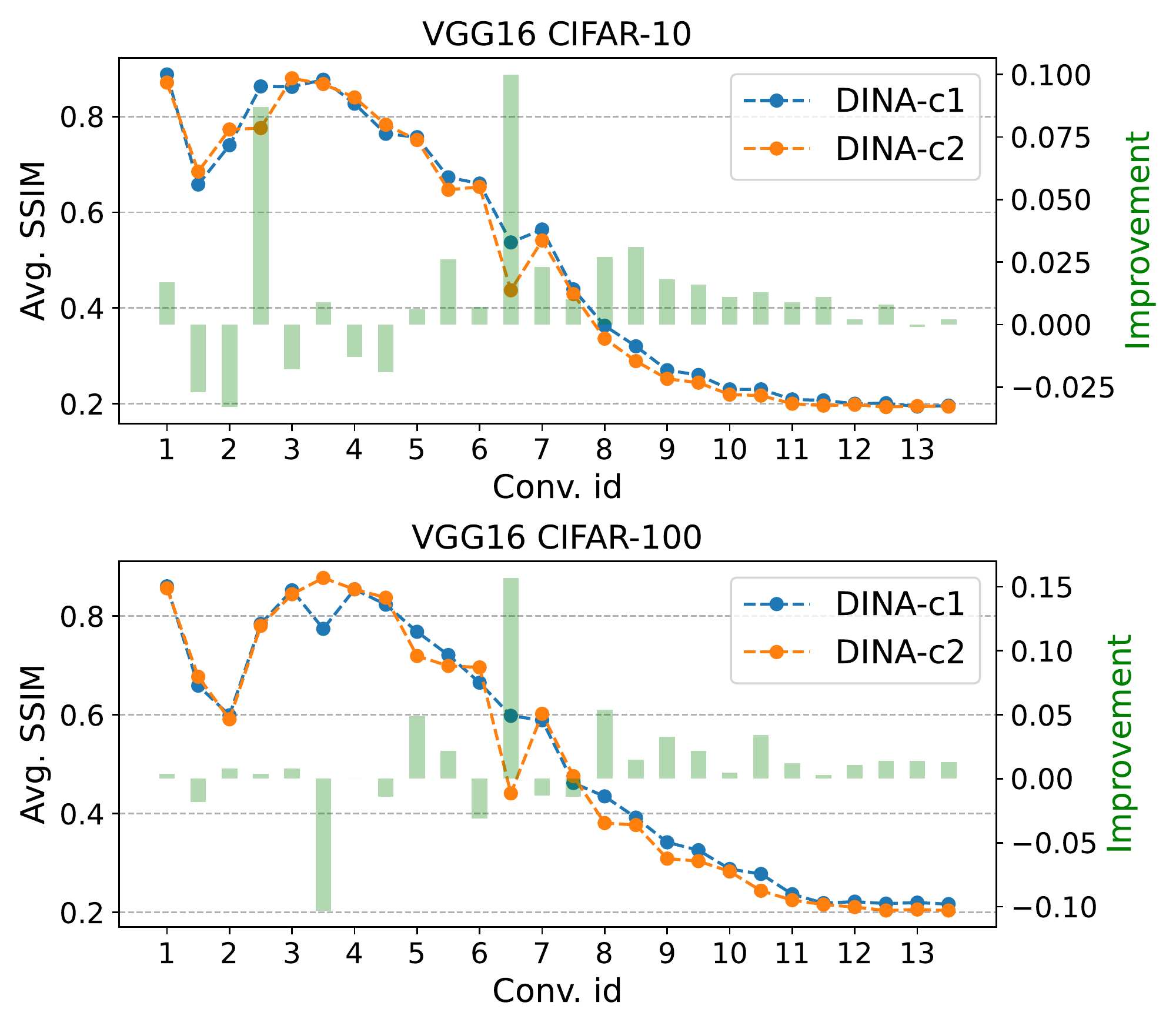}
            \vspace{-5mm}
    \caption{Attack results of DINA-c1 and DINA-c2 on VGG16. The improvements are the increased average SSIM gained by DINA-c1.}
    \label{fig:exp_alpha}
        \vspace{-5mm}
\end{figure}

\begin{figure}[t]
    \centering
    \includegraphics[width=0.85\columnwidth]{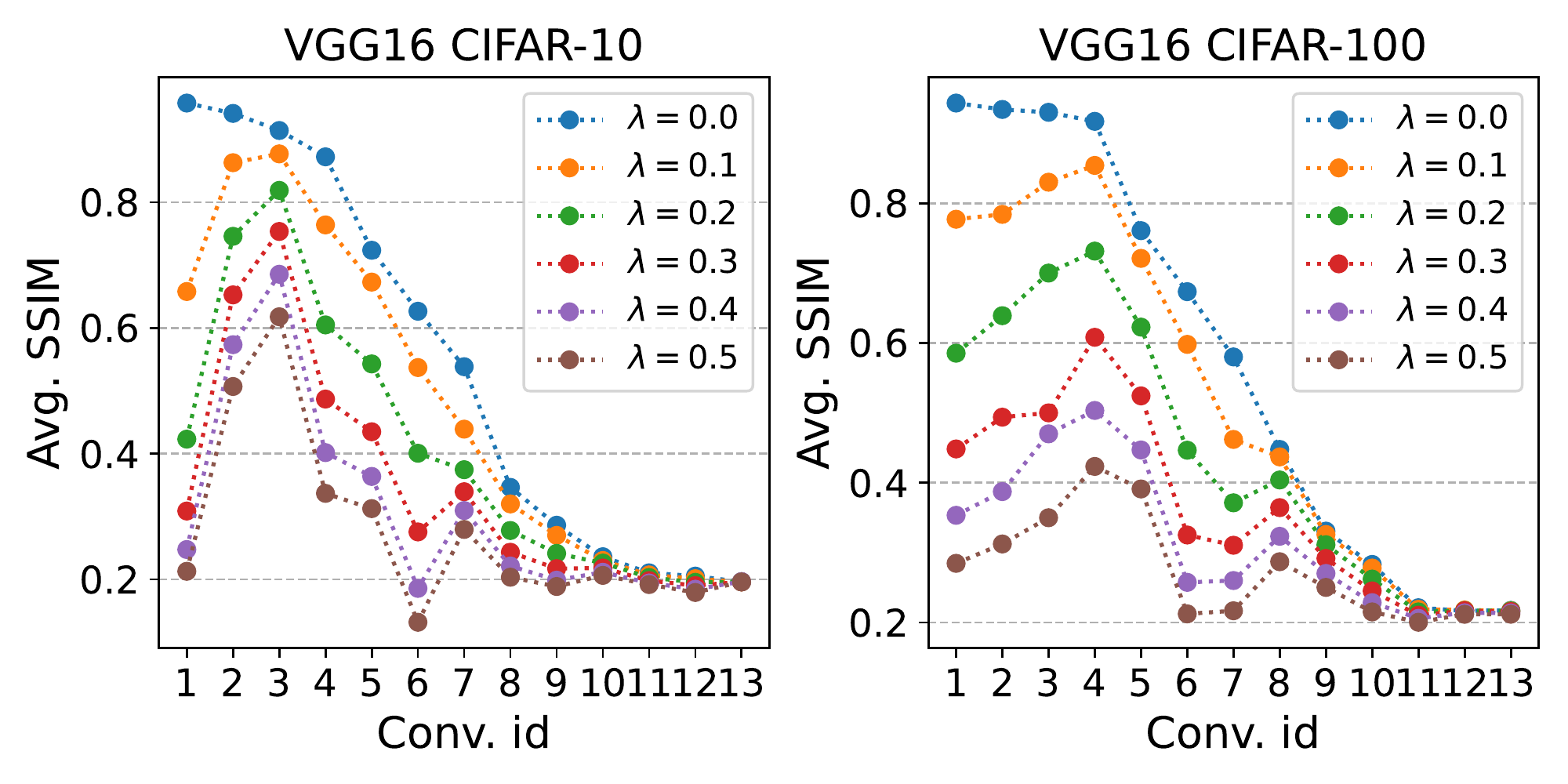}
        \vspace{-5mm}
    \caption{Adding noise as a defense against DINA.}
    \label{fig:exp_results_noise_attack}
    \vspace{-4mm}
\end{figure}

\begin{figure}[t]
    \centering
    \includegraphics[width=0.82\columnwidth]{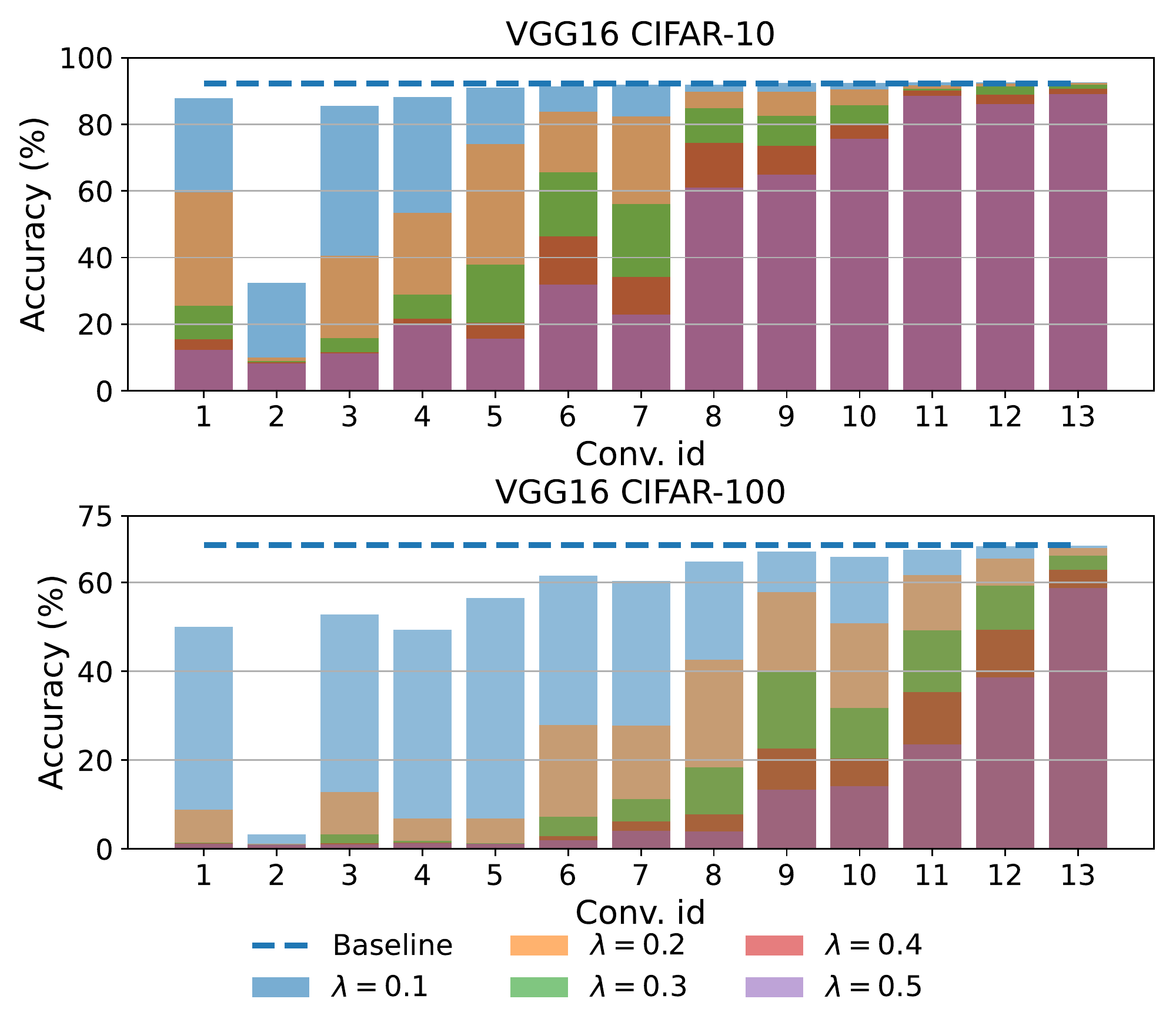}
            \vspace{-5mm}
    \caption{The effect of adding noise on accuracy on CIFAR-10 and CIFAR-100.}
    \label{fig:exp_results_noise_accuracy}
    \vspace{-5mm}
\end{figure}

\begin{figure*}[t]
    \centering
    \includegraphics[width=0.86\textwidth]{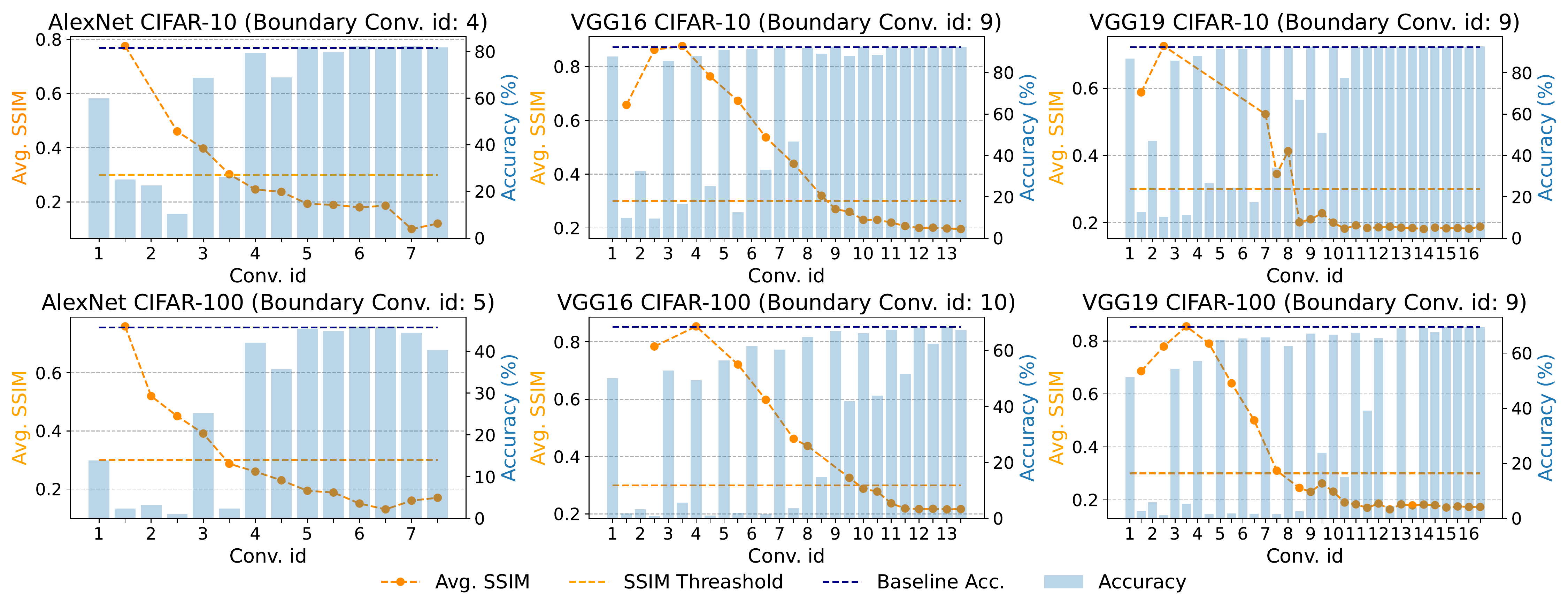}
    \vspace{-4mm}
    \caption{Find the boundary for multiple models using DINA with $0.3$ of failure threshold. Step 1: find a potential boundary after which the average SSIM begins to be below the threshold. Step 2: check the corresponding accuracy. Push the boundary at later layer %afterwards 
    until we obtain a satisfactory accuracy.}
    \vspace{-3mm}
    \label{fig:exp_results_boundary}
\end{figure*}

\subsection{Choice of DINA's Loss Coefficients}
In DINA, we use monotonously increasing coefficients $\alpha_j (j\geq0)$ in the loss function for more effective guidance on the basic inverse blocks. In this section, we compare DINA with 
increasing coefficients $\alpha_0=1, \alpha_1=3, \alpha_j=2*\alpha_{j-1} (j\geq2)$, denoted as DINA-c1, and DINA with uniform coefficients $\alpha_j=1 (j\geq 0)$, denoted as DINA-c2. Figure~\ref{fig:exp_alpha} presents the attack results where DINA-c1 achieves a higher average SSIM. We use DINA-c1 in all of our experiments.

\subsection{Effects of Adding Noise}
Now we show that adding noise helps to thwart DINA. We use DINA to attack VGG16 on CIFAR-10 and CIFAR-100 with noise magnitude changing from $0$ to $0.5$. The attack results are presented in Figure~\ref{fig:exp_results_noise_attack} and we conclude that a higher noise magnitude leads to a more vigorous defense against DINA and potentially results in an earlier boundary layer and more computational and communication savings. However, the inference accuracy degrades with the increasing noise magnitude as shown in Figure~\ref{fig:exp_results_noise_accuracy}, where the accuracy at layer $l$ is tested after feeding noised input to this layer. To balance the trade-off between the defense level and accuracy, we choose the noise magnitude of $0.1$ in our experiments.  

%We also demonstrate the effects of noise on network accuracy. To obtain the network accuracy with the noise being added at layer $l$, we extract the layer output $\mathbf{M}_l(\mathbf{x})$ and perform the later layers with $\mathbf{M}_l(\mathbf{x})+\Delta$ as the input. The effects of noise on network accuracy are recorded in Figure~\ref{fig:exp_results_noise_accuracy} where we can see that the noise toleration increases in deeper layers. This is suitable for \tecname \  as the boundary typically appears in tail of the network. Moreover, increasing the noise level leads to a significant reduction in accuracy, as expected. Based on these results, we choose the noise magnitude of $0.1$ in our experiments. 

\begin{table}[t]
    \caption{\tecname \ boundary and accuracy}
    \label{tab:exp:boundary_accuracy}
    \centering
    \resizebox{0.9\columnwidth}{!}{
    \begin{tabular}{lcccccc}
    \toprule
    \multirow{2}{*}{Dataset} & \multirow{2}{*}{Network} & Full PI & \multicolumn{2}{c}{\tecname \ ($\sigma = 0.2$)} & \multicolumn{2}{c}{\tecname \  ($\sigma = 0.3$)} \\ \cmidrule(r){3-3} \cmidrule(r){4-5} \cmidrule(r){6-7}
    & & Baseline Acc. & Boundary  &  Acc. & Boundary  &  Acc.
    \\ \midrule
 \multirow{3}{*}{CIFAR-10} & AlexNet  & $81.56$ &$5$ & $81.97$ & $4$ & $79.32$  \\   
 
  & VGG16& $92.33$ & $13.5$ & $92.61$ & $9$ & $92.49$    \\
  &   VGG19 & $92.38$ & $11$ & $92.66$  & $9$ & $92.42$    \\
  \midrule
 \multirow{3}{*}{CIFAR-100} &  AlexNet & $45.66$ & $5$ & $45.36$  & $5$ & $45.36$    \\
 & VGG16 & $68.44$  & $13.5$ & $68.44$ & $10$ & $66.53$   \\
 &   VGG19 &  $69.54$ & $11$ & $67.3$  & $9$ & $67.06$    \\
    \bottomrule
    \end{tabular}}
        \vspace{-5mm}
\end{table}

\begin{table*}
      \caption{Performance comparison of \tecname \ and Delphi/Cheetah on CIFAR-10.}
      \centering
      \label{tab:exp_comp_costs}
  \resizebox{0.95\textwidth}{!}{%
\begin{tabular}{lcccccccccc}
\toprule
\multirow{3}{*}{Network} & \multirow{3}{*}{Method} & \multicolumn{3}{c}{Full PI} &    \multicolumn{3}{c}{\tecname \  ($\sigma=0.2$)} & \multicolumn{3}{c}{\tecname \ ($\sigma=0.3$)} \\ \cmidrule(r){3-5} \cmidrule(r){6-8} \cmidrule(r){9-11}

 &  & \multicolumn{2}{c}{Latency (s)}                     & \multirow{2}{*}{Commu. (MB)}  & \multicolumn{2}{c}{Latency (s)}                     & \multirow{2}{*}{ \begin{tabular}[c]{@{}c@{}} Commu. (MB)\\ (Improv.) \end{tabular}} & \multicolumn{2}{c}{Latency (s)}                     & \multirow{2}{*}{ \begin{tabular}[c]{@{}c@{}} Commu. (MB)\\ (Improv.) \end{tabular}} \\ \cmidrule(r){3-4} \cmidrule(r){6-7} \cmidrule(r){9-10}
   &        &  LAN     & WAN   &   &  LAN (Improv.)     & WAN (Improv.)  & &  LAN (Improv.)    & WAN  (Improv.) &                                  \\ \midrule
\multirow{2}{*}{VGG16} & Delphi~\cite{delphi2020}  & $6166.47$ &       $9966.48$    & $5163$ & $6109.47$ ($\mathbf{{\sim}1x}$) &  $9869.12$ ($\mathbf{{\sim}1x}$)   & $5163$ ($\mathbf{{\sim}1x}$)  & $2351.5$ ($\mathbf{2.62x}$)  & $2568.45$ ($\mathbf{3.88x}$) & $5143$ ($\mathbf{{\sim}1x}$)  \\
& Cheetah~\cite{huang2022cheetah}  &  $13.72$  & $25.27$  &   $179.64$ 
& $14.38$ ($\mathbf{1.19x}$) &	$25.08$ ($\mathbf{1x}$)	& $163.8$	($\mathbf{1.1x}$)
& $9.38$ ($\mathbf{1.46x}$)   & $14.76$ ($\mathbf{1.71x}$)     & $71.89$ ($\mathbf{2.5x}$)\\ \midrule
% & Muse~\cite{lehmkuhl2021muse} & $1646.55$ &	$2940.86$	& $102354.34$	&	$1623.69$ ($\mathbf{1.01x}$) & $2893.67$	($\mathbf{1.02x}$) &	$101541.58$($\mathbf{1.x}$)	&	$1514.34$	($\mathbf{1.09x}$)	 & $2757.24$	($\mathbf{1.07x}$)  &	$98273.72$	($\mathbf{1.04x}$)  \\ \midrule

\multirow{2}{*}{VGG19} & Delphi~\cite{delphi2020}  & $12780.36$ &  $13265.52$  &  $5184$   & $5510.23$ ($\mathbf{2.3x}$) & $6068.12$ ($\mathbf{2.19x}$)  & $5162$  ($\mathbf{{\sim}1x}$) & $4409.95$ ($\mathbf{2.9x}$) &    $5373.34$ ($\mathbf{2.47x}$)    &   $5143$ ($\mathbf{{\sim}1x}$)   \\
 % & &	$199.61$	& $245.74$ &	 $2538.43$	&	$155.75$	($\mathbf{1.28x}$) &	$170.9$	($\mathbf{1.44x}$) &	$2519.32$	($\mathbf{{\sim}1x}$)	 &	$122.79$	($\mathbf{1.63x}$) &	$168.51$	($\mathbf{1.46x}$) &	$2509.08$	($\mathbf{1.01x}$) \\

& Cheetah~\cite{huang2022cheetah}  &  $16.81$      &        $27.67$   &  $211.4$       & $11.89$ ($\mathbf{1.51x}$) & 18.23 ($\mathbf{1.66x}$) &	89.55	($\mathbf{2.39x}$)

&  $11.51$  ($\mathbf{1.46x}$)    &  $15.23$  ($\mathbf{1.82x}$)            &    $76.83$ ($\mathbf{2.75x}$) 
% & Muse~\cite{lehmkuhl2021muse} & $1789.54$ &	$3172.25$ &	$110486.32$	 &	$1637.81$	($\mathbf{1.09x}$) 	& $2969.99$	($\mathbf{1.07x}$)  &	$105588.72$	($\mathbf{1.05x}$) 	&	$1603.89$	($\mathbf{1.12x}$)  &	$2913.73$	($\mathbf{1.09x}$)  &	$103954.77$	 ($\mathbf{1.06x}$) 

\\
\bottomrule
\end{tabular}}
\vspace{-5mm}
\end{table*}

\subsection{Find the Crypto-Clear Boundary} 

We apply the proposed boundary searching algorithm (Algorithm~\ref{alg:search_boundary}) to AlexNet and VGG16/19 %and ResNet34 
on CIFAR-10 and CIFAR-100. In the first phase of the searching process, we find a potential boundary after which the server cannot recover client's input through DINA. We then check the accuracy with noise being added at these layers and decide the boundary as the earliest layer presenting less than $2.5\%$ reduction in accuracy, a target that is similar to prior works~\cite{%jha2021deepreduce, 
cho2022selective, kundu2023learning}. This searching procedure is presented in Figure~\ref{fig:exp_results_boundary}. As users are free to tune the DINA failure threshold ($\sigma$), we show boundaries and accuracy corresponding to $\sigma$ of $0.2$ and $0.3$ in Table~\ref{tab:exp:boundary_accuracy}.

\subsection{Computational Costs}
Our implementations are built on top of Cheetah~\cite{huang2022cheetah} and Delphi~\cite{delphi2020}. All the experiments in Table~\ref{tab:exp_comp_costs} are performed on Ubuntu 22.04 with 11GB of RAM. We run our benchmarks on two network settings, i.e., LAN and WAN. The network bandwidth and round-trip time are around $384$MBps and $0.3$ms in LAN, and $44$MBps and $40$ms in WAN~\cite{huang2022cheetah}.

In Table~\ref{tab:exp_comp_costs}, we compare the full PI costs with \tecname \ with two SSIM thresholds ($0.2$ and $0.3$) on VGG16 and VGG19. When the crypto layers are performed with Delphi, \tecname \ achieves up to $3.88\times$ and $2.9\times$ speedup for VGG16 and VGG19, respectively. When the crypto layers were performed with Cheetah, \tecname \  was about $1.1\times -1.82\times$ faster than full Cheetah, and requires ${\sim}2.5\times$ less communication. \tecname \ ($\sigma=0.2$) presents similar costs with full PI for VGG16 mainly because the boundary is quite late here.

\vspace{-1mm}
\section{Conclusion}\label{sec:conclusion}
In this paper, we propose an efficient two-party PI framework, \tecname, which leverages IDPA-based privacy evaluation to relax existing PI methods. We also propose a powerful distillation-based IDPA, DINA, that can recover higher-quality images than its alternatives. In \tecname, DINA is used to find a rigid boundary that guarantees the proposed IDPA-based privacy. Users can set the privacy level by tuning the DINA's failure threshold. Moreover, \tecname \  can be applied to any existing two-party PI scheme over semi-honest threat model, and our experimental results indicate that \tecname \ helps to reduce the computational costs of the state-of-the-art PI scheme.

Besides its significant role in \tecname, DINA also helps address the privacy issue in split learning. We recognize that emerging IDPAs may throw a shadow on current \tecname. However, we are glad to replace DINA with a more aggressive IDPA in \tecname. On the other hand, we believe that cheap but effective countermeasures will appear accordingly and fortify \tecname. Our future work includes exploring and applying more defenses against IDPA to preserve client's data privacy, and embedding \tecname \  with PI methods that go beyond the semi-honest threat model, e.g., the malicious-client threat model.
\vspace{-2mm}

% \appendix
\bibliography{bibliography}
\bibliographystyle{ieeetr}
% \section{Acknowledgments}

% \newpage
% \section{Appendix}

\end{document}